# Fluctuation-Dissipation Limits in Quantum Thermoelectric Transport

Ousi Pan, Zhiqiang Fan, Shunjie Zhang, Jie Li, Jincan Chen, Shanhe Su[+]

Department of Physics, Xiamen University, Xiamen 361005, People's Republic of China

**Abstract**: As a fundamental measure of stability in nonequilibrium thermodynamics, fluctuations provide critical insight into the performance and reliability of heat engines. In this work, we establish universal fluctuation-dissipation bounds that directly link energy-current fluctuations to both the entropy production rate and steady-state transport currents. Our results are applicable to arbitrary temperature and chemical potential gradients and hold for all steady states within the framework of quantum scattering theory. These bounds remain robust even in regimes where quantum effects break classical thermodynamic uncertainty relations. We demonstrate their validity by using boxcar transmission functions and further derive constraints on the power output from the perspective of fluctuations and dissipation, offering a unified thermodynamic guideline for the design and evaluation of nanoscale and quantum thermal devices.

---

[+] sushanhe@xmu.edu.cn

*Introduction*—Dissipation serves as a quantitative measure of thermodynamic irreversibility in both physical and biological systems [1-9], whereas fluctuations—often viewed as disruptive noise—can undermine operational stability and precision [10-14]. The intrinsic link between these two concepts is embodied by the fluctuation–dissipation theorem (FDT) [15], which originated within linear-response theory [16-18], and has since been generalized to systems far from equilibrium [19-24]. Beyond theoretical advances, the FDT has been experimentally verified in various systems, including aging colloidal glasses [25-27], toroidal optical traps [28], and ensembles of bristlebots [29]. More recently, the study of the FDT has been extended to diverse quantum settings [30-33], including quantum circuits [34-36], quantum pumps [37], and quantum fluids [38]. These theoretical advances have been complemented by experimental verification in cold atomic gases [39-43] and optical microcavities [44].

A natural and compelling extension of the FDT lies in its application to thermoelectric devices. Some researchers have, for example, derived universal bounds for current fluctuations—encompassing charge, heat, and work currents—in the linear-response regime [45], analyzed charge-current fluctuations induced purely by temperature gradients [46], and examined charge, spin, and heat fluctuations under zero-current conditions [47, 48], along with entropy fluctuations in analogous settings [49]. Nevertheless, most of the prior investigations have been concentrated on charge and heat-current fluctuations within the linear-response framework or systems driven solely by temperature or chemical potential gradient [50, 51]. Recently, fluctuation–dissipation bounds for charge current were established for arbitrary nonequilibrium conditions [52]. However, energy-current fluctuations far from equilibrium remain largely unexplored within the FDT framework.

Addressing this gap is of fundamental importance, as it directly probes the stability of energy flow in quantum systems—a critical prerequisite for reliable thermodynamic operation [53]. Beyond its conceptual significance, establishing such bounds is essential for evaluating the safety, efficiency, and reliability of nanoscale thermoelectric devices, where energy-current fluctuations carry profound practical consequences. More broadly, this work represents a crucial step toward formulating general quantum thermodynamic uncertainty relations (TURs), with far-reaching implications for the understanding of nonequilibrium quantum dynamics.

In this Letter, we first introduce and establish bounds on the excess fluctuations of the energy current. Using this foundation, we then formulate universal fluctuation–dissipation bounds for the energy current in a generic two-terminal thermoelectric system. These bounds are valid for arbitrary temperature and electrochemical potential differences within the scattering-theory framework. Under nonequilibrium conditions, they define a strict admissible region for energy-current fluctuations by providing simultaneous upper and lower bounds. As they are derived solely from the properties of the Fermi-Dirac distribution, these bounds remain applicable even when classical thermodynamic uncertainty relations break down. Finally, we apply these results to derive fundamental limits on the power output of thermoelectric heat engines.

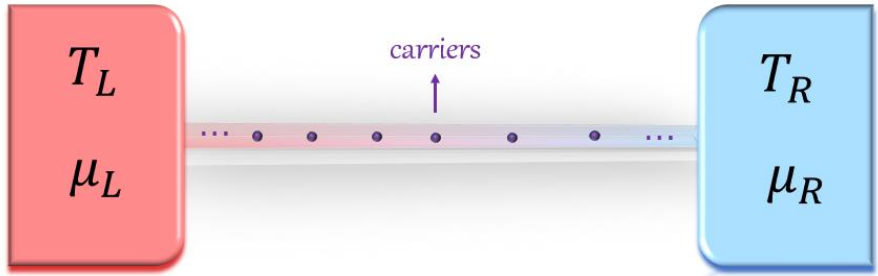

FIG. 1. The schematic diagram of a two-terminal thermoelectric system. Charge carriers and energy are exchanged between the left and right reservoirs, characterized by temperatures $T_L$ and $T_R$, and electrochemical potentials $\mu_L$ and $\mu_R$, respectively.

*Model and theory*—Our analysis is based on the Landauer-Büttiker scattering theory [54] for a two-terminal thermoelectric system operating in steady state. As depicted in Fig. 1, the temperatures of the left and right reservoirs are $T_L$ and $T_R$, and their electrochemical potentials are $\mu_L$ and $\mu_R$, respectively. Without loss of generality, we assume $T_L > T_R$. The average particle and energy currents from reservoir $L$ to reservoir $R$ are defined as [55, 56]

$$J_N = \frac{1}{h}\int_{-\infty}^{\infty} d\varepsilon D(\varepsilon)\left[f_L(\varepsilon) - f_R(\varepsilon)\right], \tag{1}$$

$$J_E = \frac{1}{h}\int_{-\infty}^{\infty} d\varepsilon \varepsilon D(\varepsilon)\left[f_L(\varepsilon) - f_R(\varepsilon)\right], \tag{2}$$

where $D(\varepsilon) \in [0,1]$ is the energy-dependent transmission function, and $f_\alpha(\varepsilon) = \{1 + \exp[(\varepsilon - \mu_\alpha)/(k_B T_\alpha)]\}^{-1}$ ($\alpha = L, R$) denotes the Fermi-Dirac distribution function for carriers (electrons or holes) with energy $\varepsilon$ in reservoir $\alpha$. Here, $h$ is Planck's constant and $k_B$ is Boltzmann's constant. The net heat currents leaving reservoir $L$ and $R$ are given by

$$J_Q^L = \frac{1}{h}\int_{-\infty}^{\infty} d\varepsilon (\varepsilon - \mu_L) D(\varepsilon)\left[f_L(\varepsilon) - f_R(\varepsilon)\right] = J_E - \mu_L J_N, \tag{3}$$

$$J_Q^R = \frac{1}{h}\int_{-\infty}^{\infty} d\varepsilon (\varepsilon - \mu_R) D(\varepsilon)\left[f_R(\varepsilon) - f_L(\varepsilon)\right] = \mu_R J_N - J_E. \tag{4}$$

At steady state, the entropy production rate of the system is given by

$$\sigma = -\frac{J_Q^L}{T_L} - \frac{J_Q^R}{T_R} = \left(\frac{1}{T_R} - \frac{1}{T_L}\right)(J_E - \varepsilon_0 J_N), \tag{5}$$

where the equilibrium energy $\varepsilon_0 = (\mu_R T_L - \mu_L T_R)/(T_L - T_R)$ refers to the energy at which the two Fermi–Dirac distributions intersect, i.e., $f_L(\varepsilon_0) = f_R(\varepsilon_0)$. The second law of thermodynamics dictates that $\sigma \geq 0$ [57-59]. An upper bound for $\sigma$ is derived in the Supplemental Material.

The total fluctuations can be separated as $S^K = \Theta_L^K + \Theta_R^K + S_{sh}^K$ [47, 52, 55, 60], where

$$\Theta_\alpha^K = \frac{1}{h}\int_{-\infty}^{\infty} d\varepsilon \xi_K^2 D(\varepsilon) f_\alpha(\varepsilon)\left[1 - f_\alpha(\varepsilon)\right], \tag{6}$$

$$S_{sh}^K = \frac{1}{h}\int_{-\infty}^{\infty} d\varepsilon \xi_K^2 D(\varepsilon)\left[1 - D(\varepsilon)\right]\left[f_L(\varepsilon) - f_R(\varepsilon)\right]^2. \tag{7}$$

The symbols are interpreted as follows: $K = N$ and $\xi_K = 1$ correspond to the particle-current fluctuations, $K = E$ and $\xi_K = \varepsilon$ describe the energy-current fluctuations, and $K = J_Q^L$ and $\xi_K = \varepsilon - \mu_L$ characterize the heat-current fluctuations of reservoir $L$. $\Theta_\alpha^K$ corresponds to the thermal-noise component arising from the interaction between the system and the finite-temperature reservoir, whereas $S_{\text{sh}}^K$ represents the shot-noise component generated by the carrier transport [46, 54]. When particle transfer occurs only at the equilibrium energy $\varepsilon_0$, the shot-noise contribution $S_{\text{sh}}^K$ vanishes. In this case, the total current fluctuations $S^K$ originate solely from the thermal noise.

*Main results*—To bound energy-current fluctuations, we first derive limits on the excess energy-current fluctuations

$$S^E - 2\Theta_L^E = \frac{1}{h}\int_{-\infty}^{\infty} d\varepsilon \, \varepsilon^2 D(f_L - f_R)(2f_L - 1) - \frac{1}{h}\int_{-\infty}^{\infty} d\varepsilon \, \varepsilon^2 D^2 (f_L - f_R)^2, \tag{8}$$

where, for brevity, we suppress the explicit energy dependence by writing $f_\alpha(\varepsilon)$ as $f_\alpha$ and $D(\varepsilon)$ as $D$. It is evident that

$$\frac{1}{h}\int_{-\infty}^{\infty} d\varepsilon \, \varepsilon^2 D^2 (f_L - f_R)^2 \geq 0. \tag{9}$$

The difficulty of simplifying the first term in Eq. (8) lies in $\varepsilon^2$ in the integrand. Therefore, we use the equation $\varepsilon^2 = \varepsilon(\varepsilon - \varepsilon_0) + \varepsilon_0(\varepsilon - \varepsilon_0) + \varepsilon_0^2$ to divide the term into three parts. The first part satisfies the inequality

$$\frac{1}{h}\int_{-\infty}^{\infty} d\varepsilon \, \varepsilon(\varepsilon - \varepsilon_0) D(f_L - f_R)(2f_L - 1) \leq \frac{1}{h}\int_{-\infty}^{\infty} d\varepsilon \, \mu_L(\varepsilon - \varepsilon_0) D(f_L - f_R)(2f_L - 1)$$
$$\leq \frac{|\mu_L|}{h}\int_{-\infty}^{\infty} d\varepsilon (\varepsilon - \varepsilon_0) D(f_L - f_R), \tag{10}$$

where we have used $(\varepsilon - \varepsilon_0)(f_L - f_R) \geq 0$ and $\varepsilon(2f_L - 1) \leq \mu_L(2f_L - 1) \leq |\mu_L|$. The bounded nature of the Fermi function, $f_L \in [0,1]$ has been employed in the derivation. Similarly, the second term obeys

$$\frac{1}{h}\int_{-\infty}^{\infty} d\varepsilon \, \varepsilon_0(\varepsilon - \varepsilon_0) D(f_L - f_R)(2f_L - 1) \leq \frac{|\varepsilon_0|}{h}\int_{-\infty}^{\infty} d\varepsilon (\varepsilon - \varepsilon_0) D(f_L - f_R), \tag{11}$$

with $(\varepsilon - \varepsilon_0)(f_L - f_R) \geq 0$ and $\varepsilon_0(2f_L - 1) \leq |\varepsilon_0|$ taken into account. Using the relations $(f_L - f_R)(2f_L - 1) \leq (f_L - f_R)[2f_L(\varepsilon_0) - 1]$ and $2f_L(\varepsilon_0) - 1 = \tanh[\Delta\mu/(2k_B\Delta T)]$, the third term can be bounded as follows [52]

$$\frac{1}{h}\int_{-\infty}^{\infty} d\varepsilon \, \varepsilon_0^2 D(f_L - f_R)(2f_L - 1) \leq \varepsilon_0^2 J_N \tanh\left(\frac{\Delta\mu}{2k_B\Delta T}\right), \tag{12}$$

where $\Delta T = T_L - T_R$ denotes the temperature difference and $\Delta\mu = \mu_L - \mu_R$ represents the electrochemical potential difference. By combining Eqs. (8)-(12), the upper bound of the excess energy-current fluctuations can be obtained

$$S^E - 2\Theta_L^E \leq \left(|\varepsilon_0| + |\mu_L|\right)\left(J_E - \varepsilon_0 J_N\right) + \varepsilon_0^2 J_N \tanh\left(\frac{\Delta\mu}{2k_B \Delta T}\right). \tag{13}$$

Combining this result with Eq. (5) gives

$$S^E - 2\Theta_L^E \leq \sigma \frac{T_L T_R \left(|\varepsilon_0| + |\mu_L|\right)}{\Delta T} + \varepsilon_0^2 J_N \tanh\left(\frac{\Delta\mu}{2k_B \Delta T}\right). \tag{14}$$

Equation (14) represents the first main result of this Letter. This bound can be used to derive the limits of energy-current fluctuations. When the upper bound is negative, the thermal noise $\Theta_L^E$ becomes the dominant contribution to the overall energy-current fluctuations. In this regime, lowering the temperature of the reservoir provides a viable route for suppressing their magnitude. Next, we examine the conditions under which each inequality reduces to an equality. The lower bound in Eq. (9) is saturated when the transmission function $D(\varepsilon)$ permits only carriers at the equilibrium energy $\varepsilon_0$ to pass. The first inequality in Eq. (10) becomes an equality if $D(\varepsilon)$ is restricted to carriers whose energy satisfies $\varepsilon = \varepsilon_0 = \mu_L$, which requires $\Delta\mu = 0$. The second inequality in Eq. (10) holds with equality when $\mu_L = 0$. Equality in Eq. (11) occurs when $D(\varepsilon)$ is confined to carriers with energy $\varepsilon = \varepsilon_0 = 0$. Similarly, the inequality in Eq. (12) turns into an equality only if carriers at the equilibrium energy $\varepsilon_0$ are transmitted. Consequently, Eq. (14) reaches equality under the most restrictive scenario, where the transmission function allows only carriers with energy $\varepsilon = \varepsilon_0 = \mu_L = \mu_R = 0$. Equality also holds trivially in the absence of transmission, i.e., $D(\varepsilon) = 0$ for all energies, since no transport occurs between the reservoirs. We note that the bound diverges in the absence of a temperature difference ($\Delta T = 0$). Thus, for small $\Delta T$, the inequality is not tight. For general cases, the excess energy-current fluctuations $S^E - 2\Theta_L^E$ approaches the upper bound under the combined conditions of a small electrochemical potential difference $|\Delta\mu|$, a large temperature difference $\Delta T$, and a transmission function—such as a boxcar function—that confines transport to a narrow energy window near $\varepsilon_0$.

Based on Eq. (14) and the nonnegativity of both thermal and shot-noise contributions to the energy-current fluctuations ($\Theta_\alpha^E \geq 0$ and $S_{\text{sh}}^E \geq 0$), we obtain the following lower bound

$$S^E \geq -\sigma \frac{T_L T_R (|\varepsilon_0|+|\mu_L|)}{\Delta T} - \varepsilon_0^2 J_N \tanh\left(\frac{\Delta\mu}{2k_B \Delta T}\right). \tag{15}$$

Under a narrow transmission window, both the thermal noise and shot-noise components approach zero. Equation (15) then satisfies the same condition as Eq. (14) when the inequality becomes an equality. To derive an upper bound on the energy-current fluctuations, we follow the approach introduced in Ref. [52]. The key step consists of substituting $D(\varepsilon)$ with $\tilde{D}(\varepsilon) = 1 - D(\varepsilon)$ in the definitions of all relevant physical quantities (e.g. $S^E$, $\Theta_L^E$, $\sigma$, and $J_N$) that enter Eq. (14). This leads to the upper bound

$$S^E \leq \Theta_{\text{UB}}^E + (\sigma_{\text{UB}} - \sigma)\frac{T_L T_R (|\varepsilon_0|+|\mu_L|)}{\Delta T} + \varepsilon_0^2 \left(\frac{\Delta\mu}{h} - J_N\right) \tanh\left(\frac{\Delta\mu}{2k_B \Delta T}\right), \tag{16}$$

where $\Theta_{\text{UB}}^E = \Theta_{L,\text{UB}}^E + \Theta_{R,\text{UB}}^E$ denotes the upper bound of thermal component of energy-current fluctuations and $\sigma_{\text{UB}} = \pi^2 k_B^2 (\Delta T)^2 \bar{T}/(3h T_L T_R) + (\Delta\mu)^2 \bar{T}/(h T_L T_R)$ is the upper bound of the entropy production rate. Here the average temperature is defined as $\bar{T} = (T_L + T_R)/2$. The upper bound of the thermal component of the energy-current fluctuations for a single reservoir is given by $\Theta_{\alpha,\text{UB}}^E = \pi^2 (k_B T_\alpha)^3/(3h) + \mu_\alpha^2 k_B T_\alpha / h$. Detailed derivations are provided in the Supplemental Material. Since $D(\varepsilon) = 0$ leads Eq. (14) to saturate, the corresponding $\tilde{D}(\varepsilon) = 1 - D(\varepsilon) = 0$ similarly causes Eq. (16) to saturate. Therefore, Eq. (16) reaches the upper bound when $D(\varepsilon) = 1$ for all energies. Equations (15) and (16) constitute the second principal result of this Letter. The bounds allow the estimation of energy-current fluctuations based on quantities that are more readily measurable. In addition, Ref. [61] demonstrates that a set of boxcar-shaped transmission functions can minimize the fluctuations under given average power and efficiency. However, experimentally realizing such precisely tailored transmission profiles remains highly challenging. In contrast, our derived bounds are applicable to all kinds of transmission functions. Furthermore, the proposed bounds offer a method for inferring the effective transmission function: fluctuations near the lower bound are indicative of a narrow transmission window, while those approaching the upper bound reflect a transmission function that is nearly unity at all energies. These bounds are universal. In particular, they apply to thermoelectric systems under arbitrary temperatures and electrochemical potentials, as long as the scattering-theory framework remains applicable. Our analysis is also extended to heat-current fluctuations in the Supplemental Material.

*Examples*—To illustrate our findings, we consider a concrete example based on a boxcar transmission model:

$$D_{\text{boxcar}}(\varepsilon) = \begin{cases} 1, & \text{if } \varepsilon \in [\varepsilon_1, \varepsilon_1 + \delta], \\ 0, & \text{otherwise}, \end{cases} \quad (17)$$

where $\delta$ is the energy window width. The corresponding energy-current fluctuations $S^E$ are plotted in Fig. 2 as a function of normalized temperature $\Delta T / \bar{T}$ for two transmission functions: $D_{\text{boxcar}}(\varepsilon)$ and $\tilde{D}_{\text{boxcar}}(\varepsilon) = 1 - D_{\text{boxcar}}(\varepsilon)$. For comparison, we also plot the lower bounds set by the classical thermodynamic uncertainty relation (TUR), a well-established fundamental constraint on nonequilibrium fluctuations. The TUR expresses a lower bound on current fluctuations in terms of the mean current and the entropy production rate [60, 62-65]:

$$S^E \geq S^E_{\text{TUR}} = 2\frac{k_B (J_E)^2}{\sigma}. \quad (18)$$

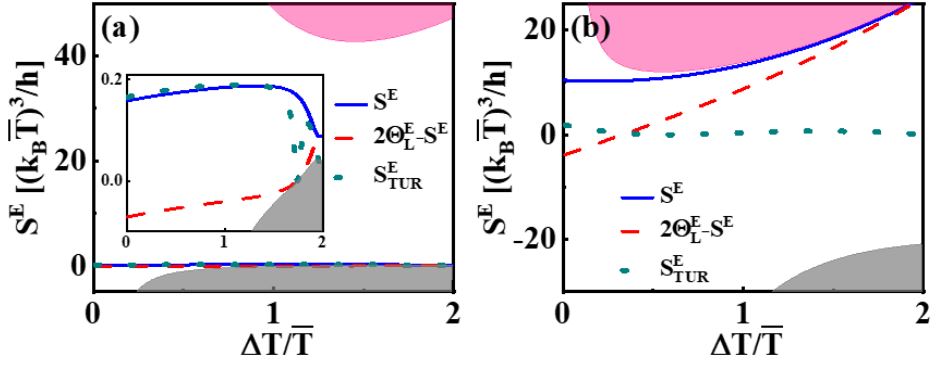

FIG. 2. Energy-current fluctuations $S^E$ as a function of the normalized temperature difference $\Delta T/\bar{T}$ for two transmission functions: (a) $D_{\text{boxcar}}(\varepsilon)$; (b) $\tilde{D}_{\text{boxcar}}(\varepsilon)=1-D_{\text{boxcar}}(\varepsilon)$. In both panels, the solid blue curves show the energy-current fluctuations $S^E$, the dashed red curves correspond to the negative excess fluctuations $2\Theta_L^E - S^E$, and the dotted dark green curves represent the classical TUR bound $S_{\text{TUR}}^E$. Fluctuations in the lower grey shaded region are theoretically inaccessible according to Eq. (15), and the upper pink shaded region corresponds to values exceeding the upper bound Eq. (16). Parameters are fixed as $\mu_L = 0$, $\mu_R = 2k_B\bar{T}$, $\delta = 0.1k_B\bar{T}$, and $\varepsilon_1 = 2.1k_B\bar{T}$. The inset in (a) provides a magnified view of the low $S^E$ region.

As shown in Fig. 2(a), at small $\Delta T/\bar{T}$, the dotted dark green curve (classical TUR bound $S_{\text{TUR}}^E$) lies above the solid blue curve (energy-current fluctuations $S^E$), indicating a slight violation of the classical TUR. This observation aligns with earlier reports that the classical TUR can be broken under certain nonequilibrium conditions. More notably, over a specific interval at larger $\Delta T/\bar{T}$, the TUR-bound curve crosses into the grey shaded region—corresponding to values that are unattainable according to our fluctuation-dissipation bound in Eq. (15).

Consistent with Eqs. (14) and (15), the energy-current fluctuations $S^E$ (solid blue curve) and the negative excess fluctuations $2\Theta_L^E - S^E$ (dashed red curve) share a common lower bound, indicated by the grey shaded region in Fig. 2. In agreement with our earlier analysis, both $S^E$ and $2\Theta_L^E - S^E$ come close to the common lower bound when transport is restricted to a narrow energy window, as illustrated by the boxcar transmission function in Fig. 2(a).

Conversely, the upper bound—marked by the pink shaded region— is realized under the condition of a fully transparent channel, i.e., $\tilde{D}(\varepsilon)=1$. Consequently, in Fig. 2(b), $S^E$ and $2\Theta_L^E - S^E$ move closer to this upper limit. In both panels, the solid blue curve lies systematically above the dashed red curve, confirming the inequality $2\Theta_L^E - S^E \leq S^E$, because $\Theta_L^E \leq S^E$.

*The bounds of power output*—We now focus specifically on the thermoelectric generator. Its power output is given by $P = J_Q^L + J_Q^R = -\Delta\mu J_N$, which requires $\Delta\mu < 0$ for positive power generation. Multiplying both sides of Eqs. (15) and (16) by $\Delta\mu$ and substituting the expression for $P$ yield corresponding bounds on the power output

$$P \leq \frac{\Delta\mu S^E}{\varepsilon_0^2 \tanh\left(\frac{\Delta\mu}{2k_B \Delta T}\right)} + \frac{\sigma \Delta\mu T_L T_R \left(|\varepsilon_0| + |\mu_L|\right)}{\varepsilon_0^2 \Delta T \tanh\left(\frac{\Delta\mu}{2k_B \Delta T}\right)}, \qquad (19)$$

$$P \geq \frac{\Delta\mu \left(S^E - \Theta_{UB}^E\right)}{\varepsilon_0^2 \tanh\left(\frac{\Delta\mu}{2k_B \Delta T}\right)} - \frac{\Delta\mu (\sigma_{UB} - \sigma) T_L T_R \left(|\varepsilon_0| + |\mu_L|\right)}{\Delta T \varepsilon_0^2 \tanh\left(\frac{\Delta\mu}{2k_B \Delta T}\right)} - \frac{(\Delta\mu)^2}{h}. \qquad (20)$$

We illustrate the bounds on power output using a double-quantum-dot transmission model [55, 66-70], described by the transmission function

$$D(\varepsilon) = \frac{\Gamma^2 \Omega^2}{\left|(\varepsilon - \varepsilon_d + i\Gamma/2)^2 - \Omega^2\right|^2}, \qquad (21)$$

where $\varepsilon_d$ denotes the energy level of the quantum dot, $\Gamma$ represents the coupling strength to the reservoirs within the wide-band approximation, and $\Omega$ signifies the tunnel coupling between the two dots.

Fig. 3(a) presents the transmission spectrum $D(\varepsilon)$ of a double quantum dot, which displays two well-separated resonant peaks. The corresponding power output $P$ plotted as the solid purple curve in Fig. 3(b), initially rises, reaches a maximum, and subsequently decreases until the system no longer functions as a heat engine. Over the entire operating range, the bound given by Eq. (19) is satisfied. Only the upper bound is plotted here, because the lower bound takes negative values and thus does not impose a physically relevant constraint for heat-engine operation.

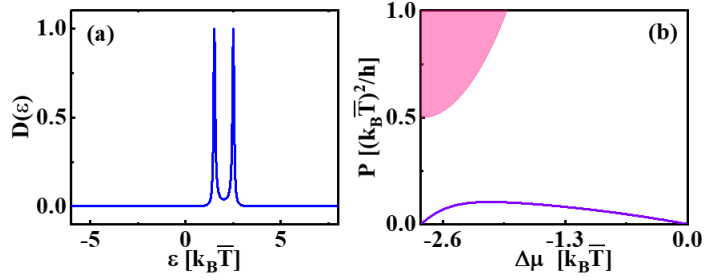

FIG. 3. (a) Transmission spectrum $D(\varepsilon)$ of a double quantum dot. (b) The power output $P$ as a function of the electrochemical potential difference $\Delta\mu$, calculated by using the transmission function shown in (a). The pink shaded regions denote values of $P$ that are unattainable given by Eq. (19). Parameters are fixed as $\Delta T = 1.8\bar{T}$, $\mu_L = -\mu_R$, $\varepsilon_d = 2k_B\bar{T}$, $\Gamma = 0.1k_B\bar{T}$, and $\Omega = 0.5k_B\bar{T}$.

*Conclusion*—This work establishes general bounds on energy-current fluctuations in two-terminal thermoelectric systems within the scattering-theory framework. These bounds are universal, holding for any transmission function and arbitrary reservoir temperatures and electrochemical potentials. They are derived by extending the previously established fluctuation-dissipation bound formalism to the energy current, and consequently apply to arbitrary nonequilibrium steady states. Crucially, they remain valid in quantum regimes where classical TURs break down. Future studies could aim to formulate TURs for arbitrary currents and to apply the present theoretical bounds to guide the design of thermoelectric heat engines with enhanced stability and performance.

**Acknowledgements**

This work has been supported by National Natural Science Foundation of China (12364008), Natural Science Foundation of Fujian Province (2023J01006), and

# Supplemental Material: Fluctuation-Dissipation Limits in Quantum Thermoelectric Transport

Ousi Pan, Zhiqiang Fan, Shunjie Zhang, Jie Li, Jincan Chen, Shanhe Su[+]

Department of Physics, Xiamen University, Xiamen 361005, People's Republic of China

This Supplemental Material provides detailed derivations of the analytical results presented in the main text, including the upper bound of the entropy production rate, the bounds on the energy-current fluctuations, and the bounds on heat-current fluctuations. For clarity, all equations and figures in this document are labeled with the prefix S [e.g., Eq. (S1) or Fig. S1]. References without this prefix refer to corresponding items in the main text [e.g., Eq. (1) or Fig. 1].

## S1. THE BOUND ON THE ENTROPY PRODUCTION RATE

In this section, we present the detailed derivations of the upper bound for the entropy production rate. Under the condition of $T_L > T_R$, the following inequalities hold:

$$\int_{-\infty}^{\varepsilon_0} d\varepsilon\, \varepsilon D(\varepsilon)\left[f_L(\varepsilon) - f_R(\varepsilon)\right] \geq \varepsilon_0 \int_{-\infty}^{\varepsilon_0} d\varepsilon\, D(\varepsilon)\left[f_L(\varepsilon) - f_R(\varepsilon)\right], \quad (S1)$$

$$\int_{\varepsilon_0}^{\infty} d\varepsilon\, \varepsilon D(\varepsilon)\left[f_L(\varepsilon) - f_R(\varepsilon)\right] \geq \varepsilon_0 \int_{\varepsilon_0}^{\infty} d\varepsilon\, D(\varepsilon)\left[f_L(\varepsilon) - f_R(\varepsilon)\right], \quad (S2)$$

where $\varepsilon_0 = (\mu_R T_L - \mu_L T_R)/(T_L - T_R)$ is recalled. We thus obtain

$$J_E \geq \varepsilon_0 J_N, \quad (S3)$$

where equality is achieved if and only if $D(\varepsilon) = 0$ for all $\varepsilon \neq \varepsilon_0$. Next, we recall that the entropy production rate is given by

$$\sigma = -\frac{J_Q^L}{T_L} - \frac{J_Q^R}{T_R} = \left(\frac{1}{T_R} - \frac{1}{T_L}\right)(J_E - \varepsilon_0 J_N). \quad (S4)$$

Substituting Eq. (S3) into Eq. (S4) directly yields $\sigma \geq 0$, thereby embodying the second

---
[+] sushanhe@xmu.edu.cn

law of thermodynamics. This conclusion aligns with the result of Ref. [1], which further extended the analysis to multi-terminal systems. Proceeding further, we insert Eq. (S3) into the definition of $J_Q^L$ provided in Eq. (3), obtaining $J_Q^L \geq PT_L/\Delta T$, where the power output of the system can be concisely expressed as $P = -\Delta\mu J_N$. Consequently, the system efficiency satisfies $\eta = P/J_Q^L \leq \Delta T/T_L = \eta_C$. Therefore, achieving the Carnot efficiency $\eta_C$ in a two-terminal thermoelectric generator requires charge carriers to be transported at a single energy $\varepsilon_0$. However, this condition simultaneously results in zero power output. This conclusion is consistent with the findings reported in Ref. [2, 3], while the derivation presented here offers a more straightforward approach.

Following the approach in Ref. [4], we replace $D(\varepsilon)$ with $\tilde{D}(\varepsilon) = 1 - D(\varepsilon)$ in Eq. (S3). Given that $\tilde{D}(\varepsilon)$ also lies in $[0,1]$, the inequality $\tilde{J}_E \geq \varepsilon_0 \tilde{J}_N$ holds, in which all physical quantities with tilde above are obtained by replacing $D(\varepsilon)$ with $\tilde{D}(\varepsilon)$ in the integrand. The inequality $\tilde{J}_E \geq \varepsilon_0 \tilde{J}_N$ becomes an equality when $D(\varepsilon) = 1$ across the entire energy spectrum. The resulting average particle current, denoted with a tilde, is

$$\tilde{J}_N = \frac{1}{h}\int_{-\infty}^{\infty} d\varepsilon \left[1 - D(\varepsilon)\right]\left[f_L(\varepsilon) - f_R(\varepsilon)\right]$$
$$= \frac{1}{h}\int_{-\infty}^{\infty} d\varepsilon \left[\theta(\mu_L - \varepsilon) - \theta(\mu_R - \varepsilon)\right] - J_N \qquad (S5)$$
$$= \frac{\Delta\mu}{h} - J_N,$$

where $\theta(x)$ is the Heaviside step function, and the relation $\int_{-\infty}^{\infty} d\varepsilon f_\alpha(\varepsilon) = \int_{-\infty}^{\infty} d\varepsilon f(\varepsilon, \mu_\alpha, T_\alpha) = \int_{-\infty}^{\infty} d\varepsilon \theta(\mu_\alpha - \varepsilon)$ has been used in the second step. The corresponding average energy current is

$$\tilde{J}_E = \frac{1}{h}\int_{-\infty}^{\infty} d\varepsilon\,\varepsilon \left[1 - D(\varepsilon)\right]\left[f_L(\varepsilon) - f_R(\varepsilon)\right]$$
$$= \frac{1}{h}\int_{-\infty}^{\infty} d\varepsilon\,\varepsilon \left[\theta(\mu_L - \varepsilon) - \theta(\mu_R - \varepsilon)\right] + A(\mu_L, T_L) - A(\mu_R, T_R) - J_E$$
$$= \frac{1}{h}\int_{\mu_R}^{\mu_L} d\varepsilon\,\varepsilon + \frac{\pi^2 k_B^2}{6h}\left(T_L^2 - T_R^2\right) - J_E \qquad (S6)$$
$$= \frac{\mu_L^2 - \mu_R^2}{2h} + \frac{\pi^2 k_B^2}{6h}\left(T_L^2 - T_R^2\right) - J_E,$$

where we have defined $A(\mu_\alpha, T_\alpha) \equiv \int_{-\infty}^{\infty} d\varepsilon\, \varepsilon [f_\alpha(\varepsilon) - \theta(\mu_\alpha - \varepsilon)]/h = (\pi k_B T_\alpha)^2/(6h)$.

Substituting Eqs. (S5) and (S6) into $\tilde{J}_E \geq \varepsilon_0 \tilde{J}_N$, we obtain the upper bound for $J_E$,

$$J_E \leq \frac{\pi^2 k_B^2 \Delta T \bar{T}}{3h} + \frac{(\Delta\mu)^2 \bar{T}}{h\Delta T} + \varepsilon_0 J_N. \tag{S7}$$

Combining Eqs. (S7) and (S4) yields an upper bound on the entropy production rate

$$\sigma \leq \frac{\pi^2 k_B^2 (\Delta T)^2 \bar{T}}{3h T_L T_R} + \frac{(\Delta\mu)^2 \bar{T}}{h T_L T_R} \equiv \sigma_{\mathrm{UB}}, \tag{S8}$$

The condition under which entropy production rate attains its theoretical limits has a clear physical meaning. When carrier transmission is restricted solely to the equilibrium energy $\varepsilon_0$, the system attains a state of thermodynamic order, thereby minimizing its entropy production rate. Conversely, permitting unrestricted energy-resolved transmission maximizes thermodynamic disorder, resulting in the upper bound of entropy production rate.

The bound is universal and applicable to thermoelectric systems operating as heat engines, refrigerators, or heat pumps. To illustrate our upper bound on the entropy production rate, we apply it to a double-quantum-dot transmission model described by Eq. (21) of the main text. The transmission functions of the double quantum dot—denoted as $D(\varepsilon)$ and $\tilde{D}(\varepsilon) = 1 - D(\varepsilon)$—are plotted in Fig. S1 (a) and (c), respectively, where the coupling strength $\Gamma$ tunes the spectral shape.

The resulting dependence of the entropy production rate $\sigma$ on the normalized temperature difference $\Delta T/\bar{T}$ is shown in Fig. S1 (b) and (d), respectively. In these figures, the pink shaded region indicates the values of $\sigma$ that are inaccessible according to Eq. (S8), the grey shaded region corresponds to values that are unattainable according to second law of thermodynamics ($\sigma \geq 0$). In Fig. S1 (a) and (b), as $\Gamma$ decreases, the transmission function $D(\varepsilon)$ becomes narrow, suppressing transport through multiple energy channels. Consequently, the calculated entropy production rate $\sigma$ approaches the theoretically derived lower bound. In contrast, Fig. S1 (c) and (d) demonstrates that as $\Gamma$ decreases, the transmission function $\tilde{D}(\varepsilon)$ approaches toward a fully transparent profile—corresponding to $D(\varepsilon)=1$ across all energies—thereby driving the entropy production rate $\sigma$ toward its upper bound. Despite considerable variations in transmission profiles, the derived bound remains universally valid across all cases examined.

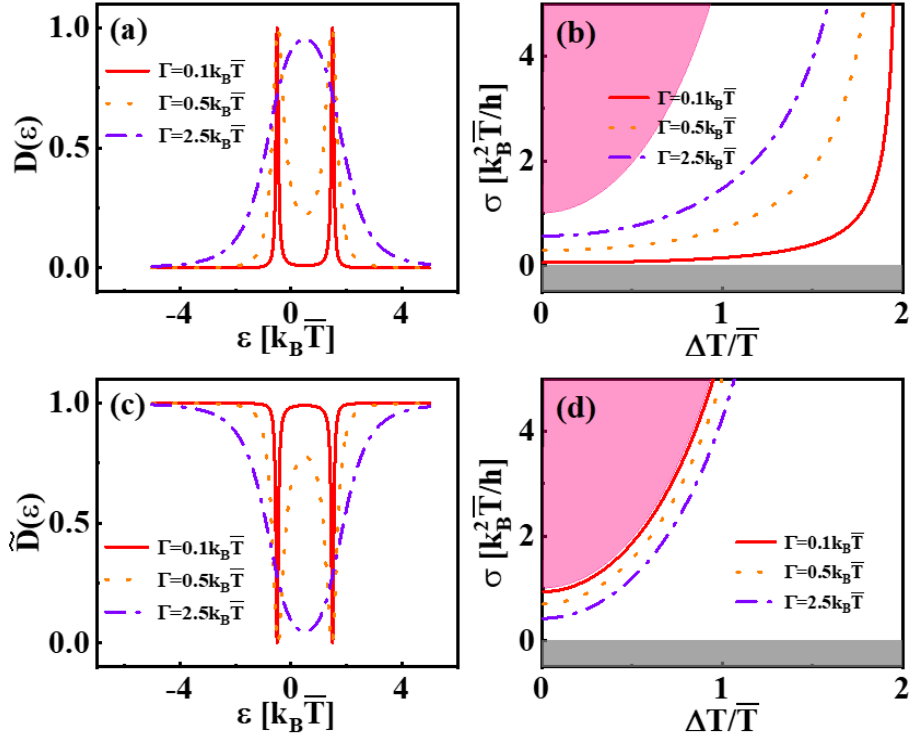

FIG. S1. (a) Transmission spectrum $D(\varepsilon)$ of a double quantum dot. (b) Corresponding entropy production rate $\sigma$ as a function of the normalized temperature difference $\Delta T/\bar{T}$. (c) Transmission spectrum $\tilde{D}(\varepsilon) = 1 - D(\varepsilon)$ of the same model. (d) Entropy production rate $\sigma$ as a function of normalized temperature difference $\Delta T/\bar{T}$, calculated using $\tilde{D}(\varepsilon)$. The upper pink shaded region indicates values that are unattainable according to Eq. (S8), while the lower grey shaded region corresponds to values $\sigma < 0$. Model parameters are fixed at $\varepsilon_d = 0.5 k_B \bar{T}$, $\Omega = k_B \bar{T}$, $\mu_L = 0$, and $\mu_R = k_B \bar{T}$. The coupling strength $\Gamma$ takes three values: $0.1 k_B \bar{T}$ (solid red), $0.5 k_B \bar{T}$ (dotted orange), and $2.5 k_B \bar{T}$ (dash-dotted purple).

## S2. FLUCTUATION-DISSIPATION BOUNDS ON ENERGY CURRENT: PROOFS OF EQS. (15) and (16)

This section is devoted to establishing fluctuation-dissipation bounds on energy current. To establish bounds for energy-current fluctuations, we first determine an upper bound for the thermal component of energy-current fluctuations. Following the approach in Sec. S1, we substitute $D$ with $\tilde{D} = 1 - D$ to obtain the thermal component of energy-current fluctuations with tilde, i.e.,

$$\begin{aligned}
\tilde{\Theta}_\alpha^E &= \frac{1}{h} \int_{-\infty}^{\infty} d\varepsilon \, \varepsilon^2 f_\alpha (1 - f_\alpha) - \Theta_\alpha^E \\
&= \frac{(k_B T_\alpha)^3}{h} \int_{-\infty}^{\infty} dx \left( x + \frac{\mu_\alpha}{k_B T_\alpha} \right)^2 \frac{e^x}{(1+e^x)^2} - \Theta_\alpha^E \\
&= \frac{(k_B T_\alpha)^3}{h} \int_{-\infty}^{\infty} dx \left[ x^2 + \left( \frac{\mu_\alpha}{k_B T_\alpha} \right)^2 \right] \frac{e^x}{(1+e^x)^2} - \Theta_\alpha^E \\
&= \frac{(k_B T_\alpha)^3}{h} \left[ \frac{\pi^2}{3} + \left( \frac{\mu_\alpha}{k_B T_\alpha} \right)^2 \right] - \Theta_\alpha^E \\
&= \Theta_{\alpha,\text{UB}}^E - \Theta_\alpha^E,
\end{aligned} \tag{S9}$$

Here we change the integration variable from $\varepsilon$ to $x = (\varepsilon - \mu_\alpha)/(k_B T_\alpha)$, such that $d\varepsilon = k_B T_\alpha dx$ in the second step. Subsequent steps make use of the even symmetry of the function $e^x/(1+e^x)^2$, followed by evaluation of the standard integrals $\int_{-\infty}^{\infty} dx\, x^2 e^x/(1+e^x)^2 = \pi^2/3$ and $\int_{-\infty}^{\infty} dx\, e^x/(1+e^x)^2 = 1$. We define $\Theta_{\text{UB}}^E = \Theta_{L,\text{UB}}^E + \Theta_{R,\text{UB}}^E$ and $\Theta_{\alpha,\text{UB}}^E = \pi^2 (k_B T_\alpha)^3 / (3h) + \mu_\alpha^2 k_B T_\alpha / h$.

Starting from Eq. (14) and using the relation $S^E = \Theta_L^E + \Theta_R^E + S_{\text{sh}}^E$, we derive

$$-S^E + 2S_{\text{sh}}^E + 2\Theta_R^E \leq \sigma \frac{T_L T_R (|\varepsilon_0| + |\mu_L|)}{\Delta T} + \varepsilon_0^2 J_N \tanh\left( \frac{\Delta \mu}{2 k_B \Delta T} \right). \tag{S10}$$

Together with the conditions $S_{\text{sh}}^E \geq 0$ and $\Theta_R^E \geq 0$, this inequality reduces to Eq. (15), representing the lower bound on energy-current fluctuations. Furthermore, Eq. (14) guarantees the validity of the inequality

$$\tilde{\Theta}_L^E - \tilde{\Theta}_R^E - \tilde{S}_{\text{sh}}^E \geq -\tilde{\sigma} \frac{T_L T_R (|\varepsilon_0| + |\mu_L|)}{\Delta T} - \varepsilon_0^2 \tilde{J}_N \tanh\left( \frac{\Delta \mu}{2 k_B \Delta T} \right). \tag{S11}$$

Note that $\tilde{S}^E_{sh} = S^E_{sh}$, and $\tilde{\sigma} = (1/T_R - 1/T_L)(\tilde{J}_E - \varepsilon_0 \tilde{J}_N)$. Combining Eqs. (S5), (S8), (S9), and (S11) with the constraint $0 \leq \Theta^E_R \leq \Theta^E_{R,\text{UB}}$, we finally obtain Eq. (16) in the main text.

## S3. FLUCTUATION-DISSIPATION BOUNDS ON HEAT CURRENT

We now turn to the heat-current fluctuations of the hot reservoir. Following a procedure analogous to that used for the energy-current fluctuations, we first examine the excess heat-current fluctuations

$$S^{J^L_Q} - 2\Theta^{J^L_Q}_L = \frac{1}{h}\int_{-\infty}^{\infty} d\varepsilon (\varepsilon - \mu_L)^2 D(f_L - f_R)(2f_L - 1) - \frac{1}{h}\int_{-\infty}^{\infty} d\varepsilon (\varepsilon - \mu_L)^2 D^2 (f_L - f_R)^2. \quad (S12)$$

To apply our earlier results to heat-current fluctuations, we use the relation $(\varepsilon - \mu_L)^2 = \varepsilon^2 - 2(\varepsilon - \varepsilon_0)\mu_L - 2\varepsilon_0 \mu_L + \mu_L^2$. The only undetermined term in the integral satisfies the bound

$$\frac{2}{h}\int_{-\infty}^{\infty} d\varepsilon (\varepsilon - \varepsilon_0)\mu_L D(f_L - f_R)(2f_L - 1) \geq -2|\mu_L|(J_E - \varepsilon_0 J_N), \quad (S13)$$

where we have used $(\varepsilon - \varepsilon_0)(f_L - f_R) \geq 0$ and $\mu_L(2f_L - 1) \geq -|\mu_L|$. Under the condition $\mu_L \varepsilon_0 \leq 0$, we can directly derive the upper bound

$$S^{J^L_Q} - 2\Theta^{J^L_Q}_L \leq \sigma \frac{T_L T_R (|\varepsilon_0| + 3|\mu_L|)}{\Delta T} + (\varepsilon_0 - \mu_L)^2 J_N \tanh\left(\frac{\Delta \mu}{2 k_B \Delta T}\right). \quad (S14)$$

We then derive the thermal component of the heat-current fluctuations with tilde:

$$\begin{aligned}
\tilde{\Theta}^{J^L_Q}_L &= \frac{1}{h}\int_{-\infty}^{\infty} d\varepsilon (\varepsilon - \mu_L)^2 f_L(1 - f_L) - \Theta^{J^L_Q}_L \\
&= \frac{\pi^2 (k_B T_L)^3}{3h} - \Theta^{J^L_Q}_L \\
&= \Theta^{J^L_Q}_{L,\text{UB}} - \Theta^{J^L_Q}_\alpha,
\end{aligned} \quad (S15)$$

$$\begin{aligned}
\tilde{\Theta}^{J^L_Q}_R &= \frac{1}{h}\int_{-\infty}^{\infty} d\varepsilon (\varepsilon - \mu_L)^2 f_R(1 - f_R) - \Theta^{J^L_Q}_R \\
&= \frac{\pi^2 (k_B T_R)^3}{3h} + \frac{k_B T_R (\mu_L - \mu_R)^2}{h} - \Theta^{J^L_Q}_R \\
&= \Theta^{J^L_Q}_{R,\text{UB}} - \Theta^{J^L_Q}_\alpha,
\end{aligned} \quad (S16)$$

where we have defined $\Theta^{J^L_Q}_{L,\text{UB}} = \pi^2 (k_B T_L)^3 / (3h)$, $\Theta^{J^L_Q}_{R,\text{UB}} = \pi^2 (k_B T_R)^3 / (3h) + k_B T_R (\Delta \mu)^2 / h$. Following an analogous procedure to that used for energy-current fluctuations, the lower and upper bounds on the heat-current fluctuations are established:

$$S^{J_Q^L} \geq -\sigma \frac{T_L T_R \left(|\varepsilon_0| + 3|\mu_L|\right)}{\Delta T} - \left(\varepsilon_0 - \mu_L\right)^2 J_N \tanh\left(\frac{\Delta \mu}{2k_B \Delta T}\right), \quad (S17)$$

$$S^{J_Q^L} \leq \Theta_{\text{UB}}^{J_Q^L} + \left(\sigma_{\text{UB}} - \sigma\right) \frac{T_L T_R \left(|\varepsilon_0| + 3|\mu_L|\right)}{\Delta T} + \left(\varepsilon_0 - \mu_L\right)^2 \left(\frac{\Delta \mu}{h} - J_N\right) \tanh\left(\frac{\Delta \mu}{2k_B \Delta T}\right), \quad (S18)$$

where we have used $\Theta_{\text{UB}}^{J_Q^L} = \Theta_{L,\text{UB}}^{J_Q^L} + \Theta_{R,\text{UB}}^{J_Q^L}$. It should be emphasized that our bounds on heat-current fluctuations are subject to the condition $\mu_L \varepsilon_0 \leq 0$.